\documentclass[12pt]{article}
\usepackage{graphicx}

\begin{document}

\title{The Leggett-Garg inequalities and the relative entropy of coherence in the Bixon-Jortner model}

\author{Hiroo Azuma${}^{1,}$\thanks{Email: hiroo.azuma@m3.dion.ne.jp}
\ \ 
and
\ \ 
Masashi Ban${}^{2,}$\thanks{Email: m.ban@phys.ocha.ac.jp}
\\
\\
{\small ${}^{1}$Advanced Algorithm \& Systems Co., Ltd.,}\\
{\small 7F Ebisu-IS Building, 1-13-6 Ebisu, Shibuya-ku, Tokyo 150-0013, Japan}\\
{\small ${}^{2}$Graduate School of Humanities and Sciences, Ochanomizu University,}\\
{\small 2-1-1 Ohtsuka, Bunkyo-ku, Tokyo 112-8610, Japan}
}

\date{\today}

\maketitle

\begin{abstract}
We investigate the Leggett-Garg inequalities and the relative entropy of coherence in the Bixon-Jortner model.
First,
we analytically derive the general solution of the Bixon-Jortner model by a technique of the Laplace transform.
So far, only a special solution has been known for this model.
The model has a single state coupled to equally spaced quasi-continuum states.
These couplings cause discontinuities in the time evolution of the occupation probability of each state.
Second, using the analytical solution,
we show that the probability distribution of the quasi-continuum states approaches the Lorentzian function
in a period of time between the initial time and the first discontinuity.
Third,
we examine violation of the Leggett-Garg inequalities and temporal variation of the relative entropy of coherence in the model.
We prove that both the inequalities and the relative entropy are invariant under transformations
of the energy-level detuning of the single state.
\end{abstract}

\section{\label{section-introduction}Introduction}
Since quantum mechanics was developed in the 1920s,
researchers have discovered various exactly soluble models and studied their properties eagerly.
However, many difficult problems still remain unsolved.
One of such problems is the Bixon-Jortner model \cite{Bixon1968}.
Because the system of the Bixon-Jortner model lies on a countably infinite dimensional Hilbert space,
its solution can be complicated and include discontinuities.
The countably infinite dimensional ket vectors are not given by the bosonic Fock space.
Moreover, energy levels of these ket vectors vary from a negative infinite value to a positive infinite value.
These facts make the problem complex.
In particular, a test for macroscopic quantum coherence has not been examined in the Bixon-Jortner model.

In this paper,
we examine the Leggett-Garg inequalities and the relative entropy of coherence in the Bixon-Jortner model.
First of all,
we give brief reviews of the topics that the current paper treats of in the following.

The Bixon-Jortner model was originally proposed in 1968 to describe intramolecular radiationless transitions
in an isolated molecule \cite{Bixon1968}.
It causes electronic relaxation between different electronic states in a polyatomic molecule,
for example,
internal conversion and intersystem crossing.
As an application of this model,
non-radiative decay processes in large molecules were discussed comprehensively in Reference~\cite{Englman1970}.
A similar model for electron transfer between biological molecules was considered in Reference~\cite{Jortner1976}.

In the Bixon-Jortner model,
a single quantum state is coupled to infinitely many other states forming a quasi-continuum,
that is to say, equally spaced background states.
These background states are not directly coupled to each other.
The time evolution of the population of the single state and a superposition of the quasi-continuum levels were examined
in References~\cite{Stey1972,Lefebvre1974,Yeh1982,Milonni1983}.
We can observe discontinuities of the step functions, or kicks, in the time evolution of the occupation probability of the single state.
These kicks are due to the couplings between the single state and the quasi-continuum background states.
The model is discussed often in the field of quantum optics
because its Hamiltonian can be diagonalized exactly
\cite{Skinner1981,Bar-Ad1997,Santra2017}.

Some researchers have tried to extend the Bixon-Jortner model from a theoretical point of view.
In References~\cite{Radmore1987,Tarzi1988,Tarzi1989},
the single state is coupled not only to the equally spaced quasi-continuum of levels but also to a true continuum.
Fano's method is used in order to analyze this extended system.
In Reference~\cite{Radmore1995},
the single state is coupled to an infinite continuum of levels with a periodic structure.
In these models,
dynamics of the single-state population has the kicks too.

The Hamiltonian of the Bixon-Jortner model is given as follows \cite{Barnett1997}:
\begin{equation}
\hat{H}
=
\hbar\Delta_{g}|g\rangle\langle g|
+
\hbar\sum_{n=-\infty}^{\infty}n\Delta|n\rangle\langle n|
+
\hbar W\sum_{n=-\infty}^{\infty}(|n\rangle\langle g|+|g\rangle\langle n|),
\label{Bixon-Jortner-Hamiltonian-0}
\end{equation}
where $|g\rangle$ represents the single state
and
$\{|n\rangle:n\in\{0, \pm 1, \pm 2, ...\}\}$ denote infinitely many other states forming the quasi-continuum.
We will put $\hbar=1$ hereafter.
In general, a wave function is given by
\begin{equation}
|\psi(t)\rangle
=
b(t)|g\rangle
+
\sum_{n=-\infty}^{\infty}
c_{n}(t)|n\rangle.
\label{wave-function-0}
\end{equation}

Reference~\cite{Barnett1997} has shown the following fact already.
If we set the initial state with
$b(0)=1$,
$c_{n}(0)=0$ for $n=0, \pm 1, \pm 2, ...$,
and
$\Delta_{g}=0$,
we can write down $b(t)$ as
\begin{equation}
b(T)
=
\exp(-\beta T)
\Biggl(
1+2\beta\sum_{k=1}^{\infty}\exp(\beta k)\frac{T-k}{k}H(T-k)L_{k}^{(1)}[2\beta(T-k)]
\Biggr),
\label{special-solution-0}
\end{equation}
where
$T=\Delta t/(2\pi)$,
$\beta=2\pi^{2}W^{2}/\Delta^{2}$,
and
$H(x)$ represents the Heaviside step function.
Here, $L_{k}^{(1)}(x)$ stands for the associated Laguerre polynomial,
whose explicit form is given by
\begin{equation}
L_{k}^{(1)}(x)
=
\frac{d}{dx}L_{k}(x),
\label{Laguerre-polynomials-0}
\end{equation}
\begin{equation}
L_{k}(x)
=
\sum_{m=0}^{k}\frac{(-x)^{m}}{(m!)^{2}}\frac{k!}{(k-m)!},
\label{associated-Laguerre-polynomials-0}
\end{equation}
where
$L_{k}(x)$ denotes the Laguerre polynomial.

The Leggett-Garg inequalities are criteria that macroscopic classical dynamics has to obey \cite{Leggett1985,Emary2014}.
They are made out of two-time correlation functions of a single system.
If the system exhibits time development in a quantum mechanical manner,
it may violate the inequalities.
We can regard the Leggett-Garg inequalities as temporal analogues of the spatial Bell's inequalities.
The Leggett-Garg inequalities have recently attracted great attention of researchers because of the latest experimental results
\cite{Palacios-Laloy2010,Goggin2011,Knee2012}.

The Leggett-Garg inequalities are defined as follows \cite{Leggett1985,Emary2014}.
We consider an observable $\hat{O}$ which has two eigenvalues $\pm 1$.
We let $O_{i}$ ($i=1,2$) denote the result of a measurement of $\hat{O}$ at time $t_{i}$ for a physical system of interest,
so that $O_{i}$ is equal to either $+1$ or $-1$.
We write the probability that specific outcomes $O_{1}$ and $O_{2}$ are measured at times $t_{1}$ and $t_{2}$ respectively as $P_{21}(O_{2},O_{1})$.

We define a two-time correlation function at times $t_{1}$ and $t_{2}$ as
\begin{equation}
C_{21}
=
\sum_{O_{1},O_{2}\in\{-1,+1\}}
O_{2}O_{1}P_{21}(O_{2},O_{1}).
\label{definition-C_21}
\end{equation}
We consider a quantity which is determined at times $t_{1}$, $t_{2}$, and $t_{3}$ as follows:
\begin{equation}
K_{3}=C_{21}+C_{32}-C_{31}.
\label{K3-definition-0}
\end{equation}
The Leggett-Garg inequality is given by
\begin{equation}
-3\leq K_{3}\leq 1.
\label{Leggett-Garg-inequality-1}
\end{equation}
We can write down another version of the Leggett-Garg inequality as
\begin{equation}
-3\leq K_{3}'\leq 1,
\label{Leggett-Garg-inequality-2}
\end{equation}
\begin{equation}
K_{3}'=-C_{21}-C_{32}-C_{31}.
\label{K3dash-definition-0}
\end{equation}

The relative entropy of coherence was proposed in Reference~\cite{Baumgratz2014} for quantifying coherence of an arbitrary quantum state.
Reference~\cite{Baumgratz2014} also suggested
the $l_{1}$ norm of coherence as a computable measure of coherence.
Both the relative entropy of coherence and the $l_{1}$ norm of coherence are indicators of quantumness of arbitrary systems.
However,
Zhang {\it et al}. showed that the latter one is not well-defined
in the Fock space for the infinite dimensional bosonic system \cite{Zhang2016}.
Thus, we focus on the former one in the current paper.
Friedenberger and Lutz have recently pointed out that the $l_{1}$ norm of coherence and the violation of the Leggett-Garg inequalities are
closely related to each other in a damped two-level system \cite{Friedenberger2017}.

The relative entropy of coherence is given by
\begin{equation}
C_{\mbox{\scriptsize rel.ent.}}(\hat{\rho})
=
S(\hat{\rho}_{\mbox{\scriptsize diag}})-S(\hat{\rho}),
\label{definition-relative-entropy-of-coherence-0}
\end{equation}
where $S$ represents the von Neumann entropy
and $\hat{\rho}_{\mbox{\scriptsize diag}}$ denotes the state obtained from the density operator $\hat{\rho}$ by deleting all off-diagonal elements
\cite{Baumgratz2014}.

In the present paper, first, instead of Equation~(\ref{special-solution-0}),
we analytically derive the general solution of $b(t)$ and
$c_{n}(t)$ for $n=0, \pm 1, \pm 2, ...$
with a non-zero value of $\Delta_{g}$ and for any initial state,
that is to say,
$b(0)$ and $c_{n}(0)$ for $n=0, \pm 1, \pm 2, ...$ given by arbitrary complex values.
Looking at the time evolution of the expansion coefficients,
we notice that the Heaviside step functions cause
kicks at $T=k$ for $k=1, 2, 3, ...$
not only in $b(t)$ but also in $\{c_{n}(t)\}$.
These kicks appear
because the single state $|g\rangle$ is coupled
to the quasi-continuum background states
$|n\rangle$ for $n=0, \pm 1, \pm 2, ...$.

Second,
putting $b(0)=1$ and $c_{n}(0)=0$ for $n=0, \pm 1, \pm 2, ...$,
we show that
$|b(t)|^{2}$ decreases exponentially as time passes for $0\le T\le 1$
and
$|c_{n}(t)|^{2}$ approaches the Lorentzian function
under the conditions $\Delta\to 0$ and $T\simeq 1$.

Third, we evaluate the Leggett-Garg inequalities for the rigorous solution
and discuss their violation.
We also estimate the relative entropy of coherence.
We show that the Leggett-Garg inequalities and the relative entropy of coherence are invariant
under the transformations $\Delta_{g}\to\pm\Delta_{g}+n\Delta$ for $n=0, \pm 1, \pm 2, ...$.

As shown in Equation~(\ref{special-solution-0}),
only the special solution for the Bixon-Jortner model has been known so far.
That is to say, only $b(t)$ for $\Delta_{g}=0$ with $b(0)=1$ and $c_{n}(0)=0$ for $n=0,\pm 1, \pm 2, ...$ has been derived analytically.
By contrast, in the present paper,
we find the general solution exactly,
namely $b(t)$ and $\{c_{n}(t)\}$ for the non-zero value of $\Delta_{g}$.
We can evaluate the Leggett-Garg inequalities and the relative entropy of coherence
because we obtain not only $b(t)$ but also $\{c_{n}(t)\}$ in rigorous forms.
Using the solution with the non-zero value of $\Delta_{g}$,
we can also confirm invariance of the inequalities and the relative entropy under the transformations of $\Delta_{g}$.
These are points of the current paper.

Although the Bixon-Jortner model was proposed in the 1960s and analysed comprehensively from the viewpoint of theoretical quantum optics in the 1970s and 1980s,
we never regard this model as an old one.
The Bixon-Jortner model is still on the active list in the field of chemical physics for studying intramolecular vibrational relaxation, for instance.
Thus, it is very important for researchers in the fields of quantum optics and chemical physics
to obtain the general solution of this model.
Derivation of the general solution enables us to examine the Leggett-Garg inequalities and the relative entropy of coherence.
Estimation of quantumness of a nontrivial model in this approach includes a novelty value.
These are the motivations of the present paper.

This paper is organized as follows.
In Sections~\ref{section-exact-solution-b} and \ref{section-exact-solution-c},
we exactly derive the general solution of $b(t)$ and $c_{n}(t)$ for $n=0, \pm 1, \pm 2, ...$, respectively.
In Section~\ref{section-Lorentzian},
we examine the behaviour of $|b(t)|^{2}$ and $|c_{n}(t)|^{2}$ for $n=0, \pm 1, \pm 2, ...$
in a period of time between the initial time and the first discontinuity.
In Section~\ref{section-Leggett-Garg},
we study the violation of the Leggett-Garg inequalities.
We confirm that we can observe the violation sporadically in the time evolution of the Bixon-Jortner model.
In Section~\ref{section-coherence},
we compute the relative entropy of coherence.
In Section~\ref{section-K3K3dash-relent-invariance},
we consider the invariance of the Leggett-Garg inequalities and the relative entropy of coherence
under the transformations $\Delta_{g}\to\pm\Delta_{g}+n\Delta$ for $n=0, \pm 1, \pm 2, ...$.
Section~\ref{section-discussion} gives a brief discussion.

\section{\label{section-exact-solution-b}The general solution of $b(t)$}
In this section, we analytically derive the general solution of $b(t)$ using a technique of the Laplace transform.
The Schr{\"o}dinger equation is given as follows:
\begin{equation}
i\frac{\partial}{\partial t}|\psi(t)\rangle=\hat{H}|\psi(t)\rangle.
\label{Schrodinger-equation-0}
\end{equation}
From Equations~(\ref{Bixon-Jortner-Hamiltonian-0}), (\ref{wave-function-0}), and (\ref{Schrodinger-equation-0}),
we obtain
\begin{eqnarray}
i\dot{b}(t)
&=&
\Delta_{g}b(t)+W\sum_{n=-\infty}^{\infty}c_{n}(t), \nonumber \\
i\dot{c}_{n}(t)
&=&
Wb(t)+n\Delta c_{n}(t).
\label{coefficients-differential-equations-0}
\end{eqnarray}
Defining the Laplace transform of an arbitrary function $f(t)$ as
\begin{equation}
{\cal L}\{f(t)\}
=
\bar{f}(s)
=
\int_{0}^{\infty}e^{-st}f(t)dt
\quad
\mbox{where $\mbox{Re}(s)>0$,}
\end{equation}
we obtain
\begin{equation}
i[-b(0)+s\bar{b}(s)]
=
\Delta_{g}\bar{b}(s)+W\sum_{n=-\infty}^{\infty}\bar{c}_{n}(s),
\label{coefficients-Laplace-equations-a}
\end{equation}
\begin{equation}
i[-c_{n}(0)+s\bar{c}_{n}(s)]
=
W\bar{b}(s)+n\Delta\bar{c}_{n}(s).
\label{coefficients-Laplace-equations-b}
\end{equation}
We solve Equation~(\ref{coefficients-Laplace-equations-b}) with respect to $\bar{c}_{n}(s)$,
and then insert the solution into Equation~(\ref{coefficients-Laplace-equations-a}).
We thus reach an equation for $\bar{b}(s)$ in the form,
\begin{equation}
\bar{b}(s)[s+i\Delta_{g}+\frac{\pi}{\Delta}W^{2}\coth\Biggl(\frac{\pi s}{\Delta}\Biggr)]
=
b(0)-iW\sum_{n=-\infty}^{\infty}c_{n}(0)\frac{1}{s+i\Delta n},
\label{b0-expression-1}
\end{equation}
where we use the formula,
\begin{equation}
\sum_{n=-\infty}^{\infty}\frac{1}{s+i\Delta n}
=
\frac{\pi}{\Delta}\coth\Biggl(\frac{\pi s}{\Delta}\Biggr).
\end{equation}

We next rewrite Equation~(\ref{b0-expression-1}) as
\begin{eqnarray}
\bar{b}(s)
&=&
[b(0)-iW\sum_{n=-\infty}^{\infty}c_{n}(0)\frac{1}{s+i\Delta n}]
[s+i\Delta_{g}+\frac{\pi}{\Delta}W^{2}\coth\Biggl(\frac{\pi s}{\Delta}\Biggr)]^{-1} \nonumber \\
&=&
[b(0)-iW\sum_{n=-\infty}^{\infty}c_{n}(0)\frac{1}{s+i\Delta n}] \nonumber \\
&&
\times
\sum_{m=0}^{\infty}\frac{(-2\pi W^{2}/\Delta)^{m}}{(s+i\Delta_{g}+\pi W^{2}/\Delta)^{m+1}}
\frac{\exp(-2\pi ms/\Delta)}{[1-\exp(-2\pi s/\Delta)]^{m}},
\label{b0-expression-2}
\end{eqnarray}
using the following formula:
\begin{equation}
\coth\Biggl(\frac{\pi s}{\Delta}\Biggr)
=
1+\frac{2\exp(-2\pi s/\Delta)}{1-\exp(-2\pi s/\Delta)}.
\end{equation}
Moreover,
from
\begin{equation}
\frac{\theta^{m}}{(1-\theta)^{m}}
=
\left\{
\begin{array}{ll}
1 & m=0, \\
\sum_{k=m}^{\infty}\theta^{k}(k-1)!/[(k-m)!(m-1)!] & m>0, \\
\end{array}
\right.
\end{equation}
we rewrite Equation~(\ref{b0-expression-2}) as follows:
\begin{eqnarray}
\bar{b}(s)
&=&
[b(0)-iW\sum_{n=-\infty}^{\infty}c_{n}(0)\frac{1}{s+i\Delta n}]
\frac{1}{s+i\Delta_{g}+\pi W^{2}/\Delta} \nonumber \\
&&
+
[b(0)-iW\sum_{n=-\infty}^{\infty}c_{n}(0)\frac{1}{s+i\Delta n}] \nonumber \\
&&
\times
\sum_{m=1}^{\infty}
\frac{(-2\pi W^{2}/\Delta)^{m}}{(s+i\Delta_{g}+\pi W^{2}/\Delta)^{m+1}}
\sum_{k=m}^{\infty}\frac{(k-1)!\exp(-2\pi sk/\Delta)}{(k-m)!(m-1)!}.
\label{b0-expression-3}
\end{eqnarray}

We are now in a position to derive $b(t)$ from the inverse Laplace transform of $\bar{b}(s)$
given by Equation~(\ref{b0-expression-3}).
We prepare the following formula.
We define $f_{m}(t)$ as
\begin{equation}
f_{m}(t)
=
\frac{t^{m}}{m!}\exp[-(\frac{\pi W^{2}}{\Delta}+i\Delta_{g})t].
\end{equation}
The Laplace transform of $f_{m}(t)$ is given by
\begin{equation}
\bar{f}_{m}(s)
=
\frac{1}{(s+i\Delta_{g}+\pi W^{2}/\Delta)^{m+1}}.
\end{equation}
Then,
we consider the inverse Laplace transform of the first term in Equation~(\ref{b0-expression-3}).
Defining
\begin{equation}
\bar{g}(s)=b(0)-iW\sum_{n=-\infty}^{\infty}c_{n}(0)\frac{1}{s+i\Delta n},
\end{equation}
and using
\begin{equation}
{\cal L}\{e^{-iat}H(t)\}
=\frac{1}{s+ia},
\end{equation}
we obtain
\begin{equation}
g(t)=b(0)\delta(t)-iW\sum_{n=-\infty}^{\infty}c_{n}(0)e^{-in\Delta t}H(t).
\end{equation}

The Laplace transform of the factor after the square parentheses of the first term in Equation~(\ref{b0-expression-3})
takes the form of $\bar{f}_{0}(s)$,
and hence its inverse Laplace transform is given by $f_{0}(t)$.
Because the Laplace transform of the convolution
\begin{equation}
g\ast f_{0}
=
\int_{0}^{t}g(t-\tau)f_{0}(\tau)d\tau.
\end{equation}
is given by
\begin{equation}
{\cal L}\{g\ast f_{0}\}
=
\bar{g}(s)\bar{f}_{0}(s),
\end{equation}
we arrive at the inverse Laplace transform of the first term of Equation~(\ref{b0-expression-3}) in the form,
\begin{eqnarray}
g\ast f_{0}
&=&
\int_{0}^{t}
[b(0)\delta(t-\tau)
-iW\sum_{n=-\infty}^{\infty}c_{n}(0)e^{-in\Delta(t-\tau)}H(t-\tau)] \nonumber \\
&&
\times
\exp[-(\frac{\pi W^{2}}{\Delta}+i\Delta_{g})\tau]d\tau \nonumber \\
&=&
b(0)\exp(-\kappa_{0}t)
+
iW\sum_{n=-\infty}^{\infty}c_{n}(0)e^{-in\Delta t}\kappa_{n}^{-1}
[\exp(-\kappa_{n}t)-1],
\end{eqnarray}
where
\begin{equation}
\kappa_{n}
=
\frac{\pi W^{2}}{\Delta}+i(\Delta_{g}-n\Delta).
\end{equation}

Next,
we consider the inverse Laplace transform of the second term in Equation~(\ref{b0-expression-3}).
Thanks to the formula,
\begin{equation}
{\cal L}\{H(t-a)f(t-a)\}
=
\exp(-sa)\bar{f}(s),
\label{Laplace-formula-1}
\end{equation}
we can disregard the factor $\exp(-2\pi sk/\Delta)$ in Equation~(\ref{b0-expression-3}) for the moment.
The rest takes the form of $\bar{g}(s)\bar{f}_{m}(s)$,
whose inverse Laplace transform is given by the convolution,
\begin{eqnarray}
g\ast f_{m}
&=&
\int_{0}^{t}
[b(0)\delta(t-\tau)
-iW\sum_{n=-\infty}^{\infty}c_{n}(0)e^{-in\Delta(t-\tau)}H(t-\tau)] \nonumber \\
&&
\times
\frac{\tau^{m}}{m!}
\exp(-\kappa_{0}\tau)d\tau \nonumber \\
&=&
b(0)\frac{t^{m}}{m!}
\exp(-\kappa_{0}t) \nonumber \\
&&
+
iW\sum_{n=-\infty}^{\infty}c_{n}(0)e^{-in\Delta t}
\sum_{j=0}^{m}
\frac{t^{j}}{j!}
\frac{1}{\kappa_{n}^{m+1-j}}
\exp(-\kappa_{n}t) \nonumber \\
&&
-iW\sum_{n=-\infty}^{\infty}c_{n}(0)e^{-in\Delta t}\frac{1}{\kappa_{n}^{m+1}},
\label{convolution-1}
\end{eqnarray}
where we use
\begin{equation}
\int_{0}^{t}\frac{\tau^{m}}{m!}\exp(-a\tau)d\tau
=
-\sum_{j=0}^{m}\frac{t^{j}}{j!}\frac{1}{a^{m+1-j}}\exp(-at)
+
\frac{1}{a^{m+1}}.
\end{equation}

From Equations~(\ref{Laplace-formula-1}) and (\ref{convolution-1}),
we arrive at the inverse Laplace transform of the second term in Equation~(\ref{b0-expression-3}),
\begin{eqnarray}
&&
H(t-\frac{2\pi}{\Delta}k)
\Biggl[
b(0)\frac{1}{m!}(t-\frac{2\pi}{\Delta}k)^{m}
\exp[-\kappa_{0}(t-\frac{2\pi}{\Delta}k)] \nonumber \\
&&
+
iW\sum_{n=-\infty}^{\infty}c_{n}(0)\exp[-in\Delta(t-\frac{2\pi}{\Delta}k)]
\sum_{j=0}^{m}\frac{1}{j!}(t-\frac{2\pi}{\Delta}k)^{j}\frac{1}{\kappa_{n}^{m+1-j}}
\exp[-\kappa_{n}(t-\frac{2\pi}{\Delta}k)] \nonumber \\
&&
-
iW\sum_{n=-\infty}^{\infty}c_{n}(0)\exp[-in\Delta(t-\frac{2\pi}{\Delta}k)]\frac{1}{\kappa_{n}^{m+1}}
\Biggr].
\end{eqnarray}

Putting the above results together,
we can write down $b(t)$ as follows:
\begin{eqnarray}
b(t)
&=&
b(0)\exp(-\kappa_{0}t)
+
iW\sum_{n=-\infty}^{\infty}c_{n}(0)\frac{e^{-in\Delta t}}{\kappa_{n}}[\exp(-\kappa_{n}t)-1] \nonumber \\
&&
+
\sum_{m=1}^{\infty}\sum_{k=m}^{\infty}(-\frac{2\pi W^{2}}{\Delta})^{m}
\frac{(k-1)!}{(k-m)!(m-1)!}
H(t-\frac{2\pi}{\Delta}k) \nonumber \\
&&
\times
\Biggl[
b(0)\frac{1}{m!}(t-\frac{2\pi}{\Delta}k)^{m}
\exp[-\kappa_{0}(t-\frac{2\pi}{\Delta}k)] \nonumber \\
&&
+
iW\sum_{n=-\infty}^{\infty}c_{n}(0)\exp[-in\Delta(t-\frac{2\pi}{\Delta}k)] \nonumber \\
&&
\times
\sum_{j=0}^{m}\frac{1}{j!}(t-\frac{2\pi}{\Delta}k)^{j}\frac{1}{\kappa_{n}^{m+1-j}}
\exp[-\kappa_{n}(t-\frac{2\pi}{\Delta}k)] \nonumber \\
&&
-
iW
\sum_{n=-\infty}^{\infty}c_{n}(0)\exp[-in\Delta(t-\frac{2\pi}{\Delta}k)]\frac{1}{\kappa_{n}^{m+1}}
\Biggl].
\end{eqnarray}

Here, we make use of
\begin{equation}
\sum_{j=0}^{m}\frac{1}{j!}(t-a)^{j}
=
\frac{1}{m!}e^{t-a}\Gamma(m+1,t-a),
\label{incomplete-gamma-function-0}
\end{equation}
where $\Gamma(z,a)$ is the incomplete gamma function defined as
\begin{equation}
\Gamma(z,a)=\int_{a}^{\infty}t^{z-1}e^{-t}dt.
\end{equation}
We also utilize the following formulae:
\begin{equation}
\sum_{m=1}^{\infty}\sum_{k=m}^{\infty}=\sum_{k=1}^{\infty}\sum_{m=1}^{k},
\end{equation}
\begin{equation}
\sum_{m=1}^{k}\frac{(k-1)!}{(k-m)!(m-1)!m!}[-(t-a)]^{m}
=
\frac{1}{k}(t-a)L_{k}^{(1)}(t-a),
\end{equation}
\begin{equation}
\sum_{m=1}^{k}\frac{(-x)^{m}}{(k-m)!(m-1)!}
=
-\frac{(1-x)^{k-1}x}{(k-1)!}.
\label{formula-finite-series-0}
\end{equation}

From the above equations, we reach the final form of the general solution $b(t)$ as
\begin{eqnarray}
b(T)
&=&
b(0)\exp(-\kappa_{0}\gamma T)
+
iW\sum_{n=-\infty}^{\infty}c_{n}(0)\kappa_{n}^{-1}\exp(-in2\pi T)[\exp(-\kappa_{n}\gamma T)-1] \nonumber \\
&&
+
b(0)\alpha\sum_{k=1}^{\infty}\frac{1}{k}\gamma(T-k)L_{k}^{(1)}[\alpha\gamma(T-k)]\exp[-\kappa_{0}\gamma(T-k)]H(T-k) \nonumber \\
&&
+
iW
\sum_{n=-\infty}^{\infty}c_{n}(0)
\sum_{k=1}^{\infty}\sum_{m=1}^{k}(-\alpha)^{m}\frac{(k-1)!}{(k-m)!(m-1)!m!} \nonumber \\
&&
\times
\exp[-in2\pi(T-k)]\kappa_{n}^{-m-1}\Gamma[m+1,\kappa_{n}\gamma(T-k)]H(T-k) \nonumber \\
&&
+
iW\alpha
\sum_{n=-\infty}^{\infty}c_{n}(0)
\sum_{k=1}^{\infty}(\kappa_{n}-\alpha)^{k-1}\kappa_{n}^{-k-1} \nonumber \\
&&
\times
\exp[-in2\pi(T-k)]H(T-k),
\label{explicit-form-b-1}
\end{eqnarray}
where
$T=\Delta t/(2\pi)$, $\alpha=2\pi W^{2}/\Delta$, and $\gamma=2\pi/\Delta$.
We pay attention to the fact that
Equation~(\ref{explicit-form-b-1}) holds for $T\geq 0$.
If we let $b(0)=1$, $c_{n}(0)=0$ for $n=0, \pm 1, \pm 2, ...$,
and $\Delta_{g}=0$,
Equation~(\ref{explicit-form-b-1}) reduces to the previous result in Equation~(\ref{special-solution-0}).

\begin{figure}
\begin{center}
\includegraphics[scale=1.0]{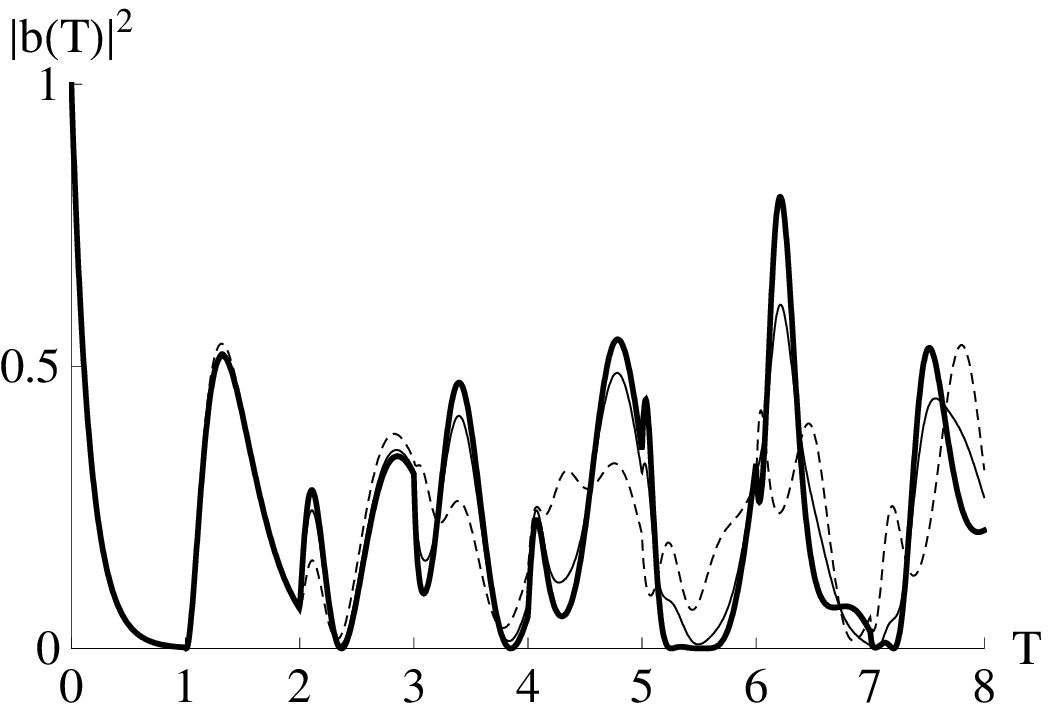}
\end{center}
\caption{Time evolution of $|b(T)|^{2}$ obtained from Equation~(\ref{explicit-form-b-1})
with $W=0.4$, $\Delta=1$, $b(0)=1$, and $c_{n}(0)=0$ for $n=0, \pm 1, \pm 2, ...$.
A thick solid curve, a thin solid curve, and a thin dashed curve represent
the cases with $\Delta_{g}=0$, $\Delta_{g}=0.12$, and $\Delta_{g}=0.24$, respectively.
The graphs have kicks at $T=1, 2, 3, ...$.
As time passes, the curves' trajectories deviate from each other depending on $\Delta_{g}$.}
\label{Figure01}
\end{figure}

\begin{figure}
\begin{center}
\includegraphics[scale=1.0]{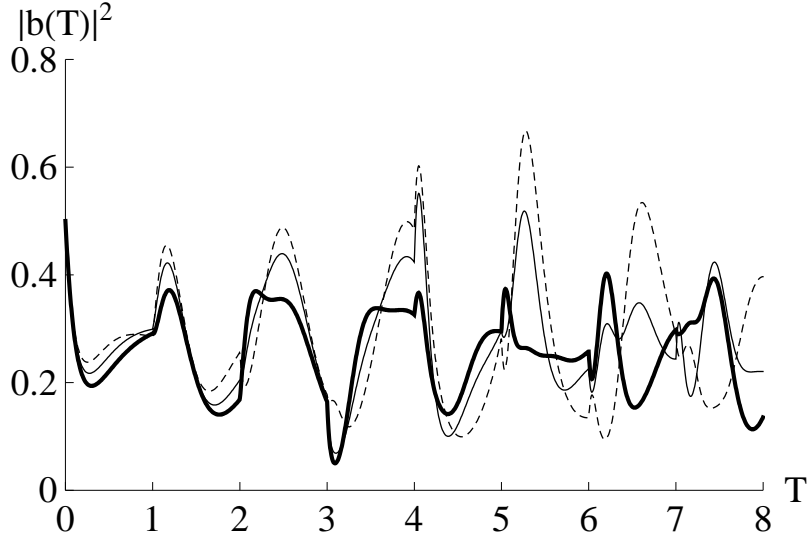}
\end{center}
\caption{Time evolution of $|b(T)|^{2}$ obtained from Equation~(\ref{explicit-form-b-1})
with $W=0.4$, $\Delta=1$, $b(0)=c_{0}(0)=1/\sqrt{2}$, and $c_{n}(0)=0$ for $n=\pm 1, \pm 2, \pm 3, ...$.
A thick solid curve, a thin solid curve, and a thin dashed curve represent
the cases with $\Delta_{g}=0$, $\Delta_{g}=0.12$, and $\Delta_{g}=0.24$, respectively.
The graphs have kicks at $T=1, 2, 3, ...$.
As time passes, the curves' trajectories deviate from each other depending on $\Delta_{g}$.}
\label{Figure02}
\end{figure}

Figure~\ref{Figure01} shows the time evolution of $|b(T)|^{2}$
obtained from Equation~(\ref{explicit-form-b-1})
with $W=0.4$, $\Delta =1$, $b(0)=1$, and $c_{n}(0)=0$ for $n=0, \pm 1, \pm 2, ...$
in the three cases,
$\Delta_{g}=0$, $\Delta_{g}=0.12$, and $\Delta_{g}=0.24$.
Figure~\ref{Figure02} shows the time evolution of $|b(T)|^{2}$
obtained from Equation~(\ref{explicit-form-b-1})
with $W=0.4$, $\Delta=1$, $b(0)=c_{0}(0)=1/\sqrt{2}$,
instead of $b(0)=1$ and $c_{0}(0)=0$,
and $c_{n}(0)=0$ for $n=\pm 1, \pm 2, \pm 3, ...$
in the three cases,
$\Delta_{g}=0$, $\Delta_{g}=0.12$, and $\Delta_{g}=0.24$.

When we compute $b(T)$ numerically according to Equation~(\ref{explicit-form-b-1}),
we replace the summation $\sum_{k=1}^{\infty}$ with $\sum_{k=1}^{8}$.
Even if we perform this simple procedure, we can obtain rigorous results of $b(T)$ in the range of $0\leq T\leq 8$.
This is because the Heaviside step function $H(T-k)$ is always included in the term where the summation of $k$ is carried out.

\section{\label{section-exact-solution-c}The general solution of $c_{n}(t)$ for
$n=0, \pm 1, \pm 2, ...$}
We have achieved the general solution of $b(t)$ in Equation~(\ref{explicit-form-b-1}).
In this section, we analytically derive the general solution of $c_{n}(t)$ for $n=0, \pm 1, \pm 2, ...$.
From Equation~(\ref{coefficients-Laplace-equations-b}), we have
\begin{equation}
\bar{c}_{n}(s)
=
-i
[W\bar{b}(s)+ic_{n}(0)]
\frac{1}{s+in\Delta}.
\end{equation}
Thus, defining
\begin{equation}
f(t)
=
Wb(t)+ic_{n}(0)\delta(t),
\end{equation}
and
\begin{equation}
g(t)
=
e^{-in\Delta t}H(t),
\end{equation}
we obtain
\begin{eqnarray}
c_{n}(t)
&=&
-i g\ast f \nonumber \\
&=&
-iW
\int_{0}^{t}
b(\tau)e^{-in\Delta(t-\tau)}d\tau
+
e^{-in\Delta t}c_{n}(0).
\label{cn-integral-1}
\end{eqnarray}

Hence,
substituting Equation~(\ref{explicit-form-b-1}) into Equation~(\ref{cn-integral-1})
and performing an integral of the convolution,
we can analytically obtain the general solution of $c_{n}(t)$.
When we carry out the convolution integral practically,
we rewrite $L_{k}^{(1)}[\alpha\gamma(T-k)]$ and $\Gamma[m+1,\kappa_{n}\gamma(T-k)]$
appearing in Equation~(\ref{explicit-form-b-1})
as finite series according to Equations~(\ref{Laguerre-polynomials-0}),
(\ref{associated-Laguerre-polynomials-0}), and (\ref{incomplete-gamma-function-0}).
Moreover, executing integrals that include the Heaviside step functions,
we make use of
\begin{equation}
\int_{0}^{t}f(\tau)H(\tau-a)d\tau
=
\int_{a}^{t}f(\tau)d\tau H(t-a).
\end{equation}

Using Equation~(\ref{formula-finite-series-0})
and the following formula:
\begin{eqnarray}
&&
\int_{a}^{t}e^{-il\Delta(\tau-a)}e^{-\kappa(\tau-a)}(\tau-a)^{j}
e^{-in\Delta(t-\tau)}d\tau \nonumber \\
&=&
e^{-in\Delta(t-a)}[\kappa+i(l-n)\Delta]^{-j-1}
[j!-\Gamma(j+1,[\kappa+i(l-n)\Delta](t-a))],
\end{eqnarray}
after complicated calculations,
we attain
\begin{eqnarray}
c_{n}(T)
&=&
-iWb(0)\tilde{\kappa}_{0,n}^{-1}[\exp(-in\Delta\gamma T)-\exp(-\kappa_{0}\gamma T)] \nonumber \\
&&
+W^{2}\sum_{l=-\infty}^{\infty}c_{l}(0)\kappa_{l}^{-1}
\Biggl[-\gamma T e^{-in\Delta\gamma T}
\frac{\exp[-i(l-n)\Delta\gamma T]-1}{-i(l-n)\Delta\gamma T} \nonumber \\
&&
+\tilde{\kappa}_{l,n}^{-1}(\exp(-in\Delta\gamma T)-\exp[-(\kappa_{l}+il\Delta)\gamma T])\Biggr] \nonumber \\
&&
+iWb(0)\alpha
\sum_{k=1}^{\infty}
\tilde{\kappa}_{0,n}^{-2}
(1-\frac{\alpha}{\tilde{\kappa}_{0,n}})^{k-1}
e^{-in\Delta\gamma T}H(T-k) \nonumber \\
&&
+iWb(0)
\sum_{k=1}^{\infty}
\sum_{m=1}^{k}
\frac{(k-1)!}{(k-m)!(m-1)!m!}(-\alpha)^{m}
\tilde{\kappa}_{0,n}^{-m-1}e^{-in\Delta\gamma T} \nonumber \\
&&
\times
\Gamma[m+1,\tilde{\kappa}_{0,n}\gamma(T-k)]H(T-k) \nonumber \\
&&
+W^{2}
\sum_{l=-\infty}^{\infty}c_{l}(0)
\sum_{k=1}^{\infty}
\sum_{m=1}^{k}
\frac{(k-1)!}{(k-m)!(m-1)!}(-\alpha)^{m}\kappa_{l}^{-m-1}
e^{-in\Delta\gamma T} \nonumber \\
&&
\times
\sum_{j=0}^{m}
\frac{1}{j!}\kappa_{l}^{j}\tilde{\kappa}_{l,n}^{-j-1}
(j!-\Gamma[j+1,\tilde{\kappa}_{l,n}\gamma(T-k)])H(T-k) \nonumber \\
&&
+W^{2}\alpha\sum_{l=-\infty}^{\infty}
c_{l}(0)\sum_{k=1}^{\infty}(\kappa_{l}-\alpha)^{k-1}\kappa_{l}^{-k-1}
\gamma(T-k)e^{-in\Delta\gamma T} \nonumber \\
&&
\times
\frac{\exp[-i\Delta(l-n)\gamma(T-k)]-1}{-i\Delta(l-n)\gamma(T-k)}H(T-k) \nonumber \\
&&
+e^{-in\Delta\gamma T}c_{n}(0),
\label{explicit-form-c-0}
\end{eqnarray}
for $n=0, \pm 1, \pm 2, ...$,
where
\begin{equation}
\tilde{\kappa}_{l,n}=\kappa_{l}+i(l-n)\Delta.
\end{equation}

Computing $c_{n}(t)$ for $n=0, \pm 1, \pm 2, ...$
numerically with Equation~(\ref{explicit-form-c-0}),
we pay attention to the following facts:
\begin{equation}
\frac{\exp[-i(l-n)\Delta\gamma T]-1}{-i(l-n)\Delta\gamma T}=1
\quad\quad
\mbox{for $l=n$},
\label{limit-calculation-1}
\end{equation}
\begin{equation}
\gamma T\frac{\exp[-i(l-n)\Delta\gamma T]-1}{-i(l-n)\Delta\gamma T}
=
\frac{\exp[-i(l-n)\Delta\gamma T]-1}{-i(l-n)\Delta}
\quad\quad
\mbox{for $l\neq n$},
\label{limit-calculation-2}
\end{equation}
\begin{equation}
\frac{\exp[-i\Delta(l-n)\gamma(T-k)]-1}{-i\Delta(l-n)\gamma(T-k)}=1
\quad\quad
\mbox{for $l=n$},
\label{limit-calculation-3}
\end{equation}
\begin{eqnarray}
\gamma(T-k)\frac{\exp[-i\Delta(l-n)\gamma(T-k)]-1}{-i\Delta(l-n)\gamma(T-k)}
&=&
\frac{\exp[-i\Delta(l-n)\gamma(T-k)]-1}{-i\Delta(l-n)} \nonumber \\
&&
\quad\quad
\mbox{for $l\neq n$}.
\label{limit-calculation-4}
\end{eqnarray}
Above equations are useful
when we avoid the division by zero in the numerical calculations of Equation~(\ref{explicit-form-c-0}) for $T=0, 1, 2, ...$.

\section{\label{section-Lorentzian}The behaviour of $|b(t)|^{2}$ for $0\le t\le 2\pi/\Delta$ and
that of $|c_{n}(t)|^{2}$ for $n=0, \pm 1, \pm 2, ...$ under the conditions $\Delta\to 0$ and $t\simeq 2\pi/\Delta$}
In this section,
putting $b(0)=1$ and $c_{n}(0)=0$ for $n=0, \pm 1, \pm 2, ...$,
we examine the behaviour of $|b(t)|^{2}$ for $0\le t\le 2\pi/\Delta$ and that of $|c_{n}(t)|^{2}$ for $n=0, \pm 1, \pm 2, ...$
under the conditions $\Delta\to 0$ and $t\simeq 2\pi/\Delta$.

Thus, we only have to investigate $b(T)$ and $c_{n}(T)$ for $n=0, \pm 1, \pm 2, ...$
in the range of $0\leq T\leq 1$,
in other words, a period of time between the initial time and the first discontinuity.
Setting $b(0)=1$ and $c_{n}(0)=0$ for $n=0, \pm 1, \pm 2, ...$,
we can rewrite Equations~(\ref{explicit-form-b-1}) and (\ref{explicit-form-c-0}) as follows:
\begin{equation}
b(T)
=
\exp(-\kappa_{0}\gamma T),
\label{simple-b-form-1}
\end{equation}
\begin{equation}
c_{n}(T)
=
-iW\tilde{\kappa}_{0,n}^{-1}
[\exp(-in\Delta\gamma T)-\exp(-\kappa_{0}\gamma T)].
\label{simple-cn-form-1}
\end{equation}

From Equation~(\ref{simple-b-form-1}),
we obtain
\begin{equation}
|b(t)|^{2}
=
\exp
(-\frac{2\pi W^{2}}{\Delta}t).
\end{equation}
Hence,
$|b(t)|^{2}$ decreases exponentially as time passes in the range of $0\le t\le 2\pi/\Delta$.

We can rewrite Equation~(\ref{simple-cn-form-1}) as
\begin{equation}
c_{n}(t)
=
-\frac{iW}{(\pi W^{2}/\Delta)+i(\Delta_{g}-n\Delta)}
[\exp(-in\Delta t)-\exp(-(\frac{\pi W^{2}}{\Delta}+i\Delta_{g})t)].
\end{equation}
Here, we impose the following two conditions simultaneously upon $c_{n}(t)$.
The first one is $\Delta\to 0$.
The second one is $T\simeq 1$, that is to say, $t\simeq 2\pi/\Delta$.
Then,
the following inequality holds:
\begin{equation}
|\exp(-in\Delta t)|
\gg
|\exp[-(\frac{\pi W^{2}}{\Delta}+i\Delta_{g})t]|,
\end{equation}
and we reach
\begin{equation}
|c_{n}(t)|^{2}
\simeq
\frac{W^{2}}{(\pi W^{2}/\Delta)^{2}+(\Delta_{g}-n\Delta)^{2}}.
\end{equation}
Hence,
we conclude that the probability distribution of
$|c_{n}(t)|^{2}$ for the variable $n$ is approximately given by the Lorentzian function
under the conditions of $\Delta\to 0$ and $t\simeq 2\pi/\Delta$.

\section{\label{section-Leggett-Garg}The violation of the Leggett-Garg inequalities}
In this section,
we examine the violation of the Leggett-Garg inequalities in the Bixon-Jortner model.

The Leggett-Garg inequalities,
which are given in Equations~(\ref{K3-definition-0}), (\ref{Leggett-Garg-inequality-1}),
(\ref{Leggett-Garg-inequality-2}), and (\ref{K3dash-definition-0}),
hold in the case where the classical notions of macroscopic realism are dominant.
However, it is possible that we observe violation of these inequalities in quantum mechanical systems.
Actually, the violation appears sporadically in the time evolution of the Bixon-Jortner model.
We explain this fact below.

First of all, we choose the following operator $\hat{O}$ as the observable that has two eigenvalues $\pm 1$:
\begin{equation}
\hat{O}=|g\rangle\langle g|-\sum_{n=-\infty}^{\infty}|n\rangle\langle n|.
\end{equation}
We define times $t_{1}$, $t_{2}$, and $t_{3}$ as
\begin{equation}
t_{1}=0,
\quad
t_{2}=\tau,
\quad
t_{3}=2\tau,
\quad
\tau\geq 0.
\label{Leggett-Garg-time-0}
\end{equation}
We put an initial state at time $t_{1}=0$ as
\begin{equation}
b(0)=1,
\quad
c_{n}(0)=0
\quad
\mbox{for $n=0, \pm 1, \pm 2, ...$}.
\end{equation}

Second, we compute $C_{21}$ defined in Equation~(\ref{definition-C_21}).
We pay attention to the fact that $O_{1}=+1$ holds with a probability of unity.
Thus, we can write down $C_{21}$ as
\begin{equation}
C_{21}
=
\sum_{O_{2}\in\{-1,+1\}}
O_{2}P_{21}(O_{2},+1).
\end{equation}
Here, we introduce the following notation.
We write the expansion coefficients of $|g\rangle$ and $\{|n\rangle:n=0, \pm 1, \pm 2, ...\}$
at time $t$
with an initial condition $b(0)$ and $\{c_{m}(0):m=0, \pm 1, \pm 2, ...\}$ at time $t=0$ as
\begin{eqnarray}
&&
b(b(0),\{c_{m}(0)\};t), \nonumber \\
&&
c_{n}(b(0),\{c_{m}(0)\};t) \quad \mbox{for $n=0, \pm 1, \pm 2, ...$}.
\end{eqnarray}
On the one hand the probability that we observe $O_{2}=+1$ by the measurement is given by
\begin{equation}
P_{21}(+1,+1)=|b(1,\{0\};\tau)|^{2}.
\end{equation}
On the other hand the probability that we obtain $O_{2}=-1$ by the measurement is given by
\begin{eqnarray}
P_{21}(-1,+1)
&=&
\sum_{n=-\infty}^{\infty}|c_{n}(1,\{0\};\tau)|^{2} \nonumber \\
&=&
1-|b(1,\{0\};\tau)|^{2}.
\end{eqnarray}
Hence, we obtain
\begin{equation}
C_{21}=2|b(1,\{0\};\tau)|^{2}-1.
\label{C21-form-1}
\end{equation}
Similarly, we reach
\begin{equation}
C_{31}=2|b(1,\{0\};2\tau)|^{2}-1.
\label{C31-form-1}
\end{equation}

Third, we compute $C_{32}$.
The probability that we observe $O_{2}=+1$ at time $t_{2}=\tau$ is given by
$|b(1,\{0\};\tau)|^{2}$.
When we obtain $O_{2}=+1$,
the wave function irreversibly reduces to $|\psi(\tau)\rangle=|g\rangle$.
Thus, the probability that we obtain $O_{2}=+1$ and $O_{3}=+1$ by the measurements is given by
\begin{equation}
P_{32}(+1,+1)
=
|b(1,\{0\};\tau)|^{4}.
\end{equation}
In contrast, the probability that we observe the outcomes of the measurements $O_{2}=+1$ and $O_{3}=-1$ is given by
\begin{eqnarray}
P_{32}(-1,+1)
&=&
|b(1,\{0\};\tau)|^{2}\sum_{n=-\infty}^{\infty}|c_{n}(1,\{0\};\tau)|^{2} \nonumber \\
&=&
|b(1,\{0\};\tau)|^{2}(1-|b(1,\{0\};\tau)|^{2}).
\end{eqnarray}

Next, the probability that we observe $O_{2}=-1$ at time $t_{2}=\tau$ is given by
\begin{equation}
\sum_{n=-\infty}^{\infty}
|c_{n}(1,\{0\};\tau)|^{2}
=
1-|b(1,\{0\};\tau)|^{2}.
\end{equation}
Then, the wave function irreversibly reduces to the following state:
\begin{equation}
|\psi(\tau)\rangle
=
\sum_{n=-\infty}^{\infty}\tilde{c}_{n}|n\rangle,
\end{equation}
\begin{equation}
\tilde{c}_{n}
=
(1-|b(1,\{0\};\tau)|^{2})^{-1/2}c_{n}(1,\{0\};\tau)
\quad
\mbox{for $n=0, \pm 1, \pm 2, ...$}.
\label{definition-c-tilde-n}
\end{equation}
On the one hand the probability that we obtain $O_{2}=-1$ and $O_{3}=+1$ by the measurements is given by
\begin{equation}
P_{32}(+1,-1)
=
(1-|b(1,\{0\};\tau)|^{2})|b(0,\{\tilde{c}_{n}\};\tau)|^{2}.
\end{equation}
On the other hand the probability that we observe the outcomes of the measurements $O_{2}=-1$ and $O_{3}=-1$ is given by
\begin{eqnarray}
P_{32}(-1,-1)
&=&
(1-|b(1,\{0\};\tau)|^{2})\sum_{n=-\infty}^{\infty}|c_{n}(0,\{\tilde{c}_{m}\};\tau)|^{2} \nonumber \\
&=&
(1-|b(1,\{0\};\tau)|^{2})(1-|b(0,\{\tilde{c}_{m}\};\tau)|^{2}).
\end{eqnarray}
Thus, we reach
\begin{equation}
C_{32}
=
|b(1,\{0\};\tau)|^{2}(2|b(1,\{0\};\tau)|^{2}-1)
+
(1-|b(1,\{0\};\tau)|^{2})(1-2|b(0,\{\tilde{c}_{m}\};\tau)|^{2}).
\label{C32-form-1}
\end{equation}
Hence, from these results,
we can compute $K_{3}$ and $K_{3}'$.

\begin{figure}
\begin{center}
\includegraphics[scale=1.0]{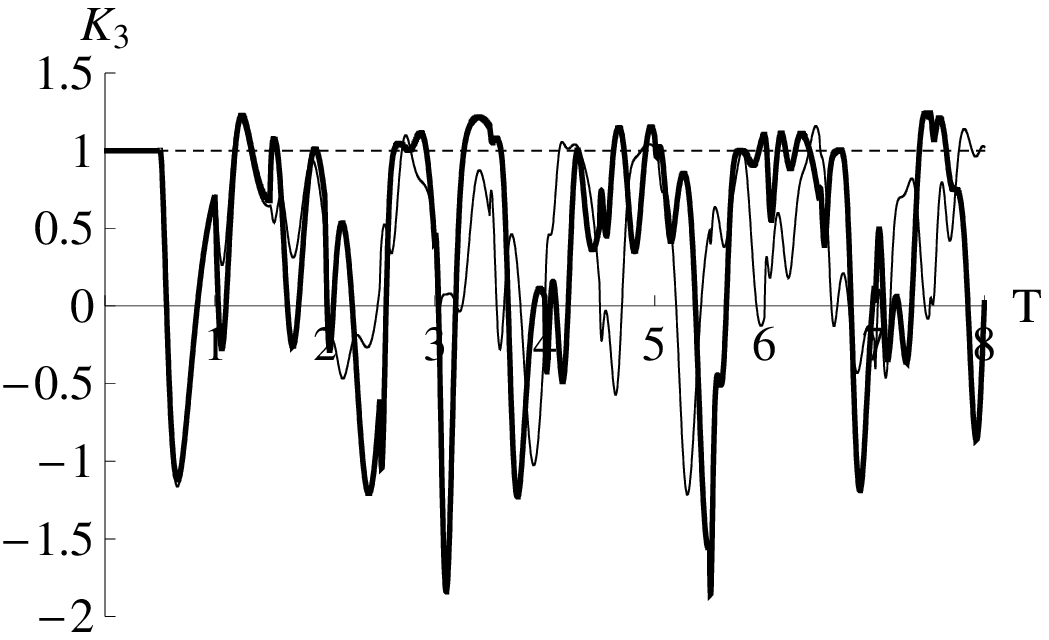}
\end{center}
\caption{Time evolution of $K_{3}$ with $W=0.4$ and $\Delta=1$ for $0\leq T\leq 8$.
A variable of the horizontal axis represents $T=\Delta\tau/(2\pi)$,
where $\tau$ is defined in Equation~(\ref{Leggett-Garg-time-0}).
A thick solid curve and a thin solid curve stand for the cases with $\Delta_{g}=0$ and $\Delta_{g}=0.24$, respectively.
For $\Delta_{g}=0$ and $0\leq T\leq 8$,
a sum of intervals of time during which $K_{3}>1$ holds is equal to $1.68$.
For $\Delta_{g}=0.24$ and $0\leq T\leq 8$,
it is equal to $0.833$.
Thus, we can consider that the quantumness of the system with $\Delta_{g}=0$ is superior to that with $\Delta_{g}=0.24$.
In these graphs, we cannot find a moment at which $K_{3}<-3$ is observed.}
\label{Figure03}
\end{figure}

\begin{figure}
\begin{center}
\includegraphics[scale=1.0]{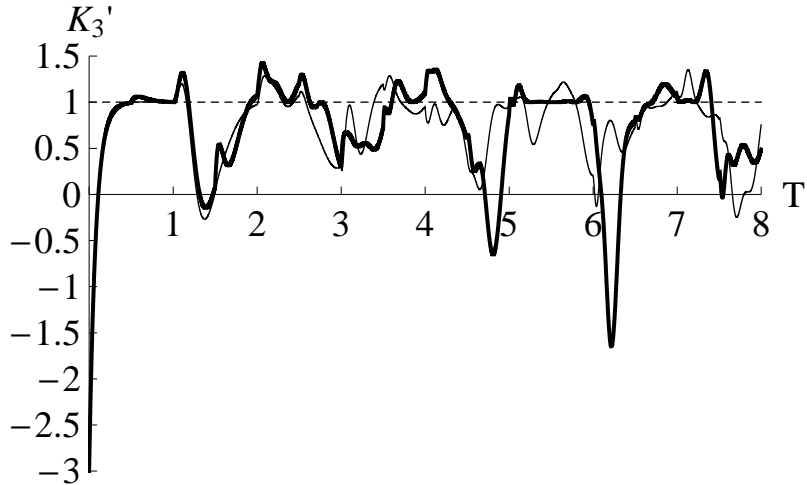}
\end{center}
\caption{Time evolution of $K_{3}'$ with $W=0.4$ and $\Delta=1$ for $0\leq T\leq 8$.
A variable of the horizontal axis represents $T=\Delta\tau/(2\pi)$,
where $\tau$ is defined in Equation~(\ref{Leggett-Garg-time-0}).
A thick solid curve and a thin solid curve stand for the cases with $\Delta_{g}=0$ and $\Delta_{g}=0.24$, respectively.
For $\Delta_{g}=0$ and $0\leq T\leq 8$,
a sum of intervals of time during which $K_{3}'>1$ holds is equal to $3.67$.
For $\Delta_{g}=0.24$ and $0\leq T\leq 8$,
it is equal to $2.17$.
Thus, we can consider that the quantumness of the system with $\Delta_{g}=0$ is superior to that with $\Delta_{g}=0.24$.
In these graphs, we cannot find a moment at which $K_{3}'<-3$ is observed.}
\label{Figure04}
\end{figure}

Figures~\ref{Figure03} and \ref{Figure04} show time evolution of $K_{3}$ and $K_{3}'$ respectively with $W=0.4$, $\Delta=1$, and $0\leq T\leq 8$
in the two cases, $\Delta_{g}=0$ and $\Delta_{g}=0.24$.
To carry out numerical calculations of $K_{3}$ and $K_{3}'$,
we use the Fortran compiler with the quadruple precision for real variables.
In actual calculations,
instead of infinite dimensional space,
we assume that the dimension of the Hilbert space is finite and its orthogonal basis is given by
$|g\rangle$ and $\{|n\rangle:n=0, \pm 1, \pm 2, ..., \pm 1000\}$.
Looking at Figures~\ref{Figure03} and \ref{Figure04},
we notice that there are periods during which either $K_{3}>1$ or $K_{3}'>1$ holds.

\section{\label{section-coherence}The relative entropy of coherence}
In this section,
we examine the relative entropy of coherence in the Bixon-Jortner model.
Putting a pure state with $b(0)=1$ and $c_{n}(0)=0$ for $n=0, \pm 1, \pm 2, ...$ at time $t=0$ as an initial state
and letting it evolve in time,
we investigate $C_{\mbox{\scriptsize rel.ent.}}$
given by Equation~(\ref{definition-relative-entropy-of-coherence-0})
as a function of the time.
We obtain
\begin{equation}
S(\hat{\rho})=S(|\psi(t)\rangle\langle\psi(t)|)=0
\quad
\forall t\geq 0,
\end{equation}
with ease.
Thus, we obtain
\begin{equation}
C_{\mbox{\scriptsize rel.ent.}}
=
-|b(1,\{0\};t)|^{2}\ln[|b(1,\{0\};t)|^{2}]
-
\sum_{n=-\infty}^{\infty}
|c_{n}(1,\{0\};t)|^{2}\ln[|c_{n}(1,\{0\};t)|^{2}].
\label{C_rel_ent_formula-1}
\end{equation}

\begin{figure}
\begin{center}
\includegraphics[scale=1.0]{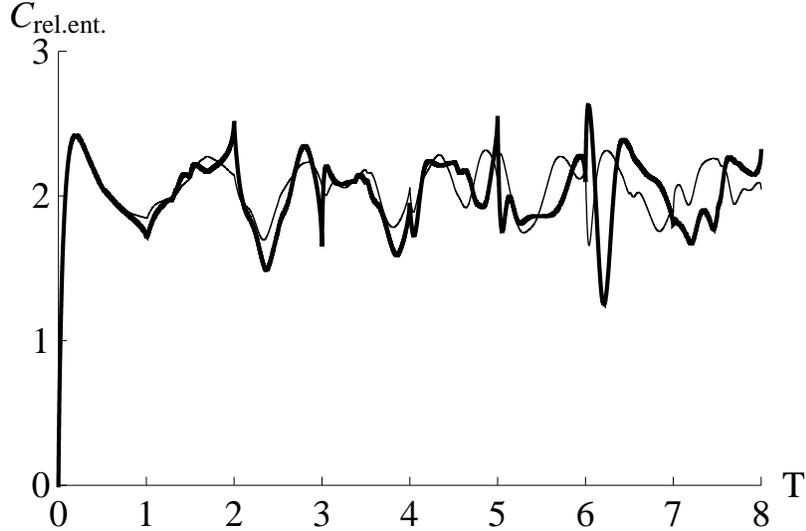}
\end{center}
\caption{Time evolution of $C_{\mbox{\scriptsize rel.ent.}}$ with $W=0.4$ and $\Delta=1$ for $0\leq T\leq 8$.
A thick solid curve and a thin solid curve represent the cases with $\Delta_{g}=0$ and $\Delta_{g}=0.24$, respectively.
For $\Delta_{g}=0$ and $0\leq T\leq 8$, an average of $C_{\mbox{\scriptsize rel.ent.}}$ is equal to $2.016$.
For $\Delta_{g}=0.24$ and $0\leq T\leq 8$, it is equal to $2.053$.
Thus, we can consider that the average of $C_{\mbox{\scriptsize rel.ent.}}$ is not sensitive to the variance of $\Delta_{g}$.
However, as time passes, we can observe that the curves' trajectories deviate from each other.}
\label{Figure05}
\end{figure}

Figure~\ref{Figure05} shows time evolution of $C_{\mbox{\scriptsize rel.ent.}}$
with $W=0.4$ and $\Delta=1$ for $0\leq T\leq 8$ in the two cases,
$\Delta_{g}=0$ and $\Delta_{g}=0.24$.
Numerical calculations are carried out as explained at the end of Section~\ref{section-Leggett-Garg}.

\section{\label{section-K3K3dash-relent-invariance}Invariance of $K_{3}$, $K_{3}'$, and $C_{\mbox{\small rel.ent.}}$
under the transformations of $\Delta_{g}\to\pm\Delta_{g}+n\Delta$ for $n=0, \pm 1, \pm 2, ...$}
In this section, we explain that $K_{3}$, $K_{3}'$, and $C_{\mbox{\scriptsize rel.ent.}}$
are invariant under the transformations of $\Delta_{g}\to\pm\Delta_{g}+n\Delta$ for $n=0, \pm 1, \pm 2, ...$
with constant values $\Delta$ and $W$.
Thus, for example,
if we regard $K_{3}$, $K_{3}'$, and $C_{\mbox{\scriptsize rel.ent.}}$ as functions of the time,
each of them for $\Delta_{g}=0.2$ is identical to that for $\Delta_{g}=0.8$ and that for $\Delta_{g}=1.2$
with constants $\Delta=1$ and $W=0.4$.

First, we prove that $K_{3}$ and $K_{3}'$ are invariant under transformations $\Delta_{g}\to\Delta_{g}+n\Delta$
for $n=0, \pm 1, \pm 2, ...$.
The transformation of $\Delta_{g}\to\Delta_{g}+n\Delta$ implies
that the Hamiltonian $\hat{H}$ given by Equation~(\ref{Bixon-Jortner-Hamiltonian-0}) is replaced with
\begin{eqnarray}
\hat{H}
&\to&
(\Delta_{g}+n\Delta)|g\rangle\langle g|
+
\sum_{m=-\infty}^{\infty}m\Delta|m\rangle\langle m|
+
W\sum_{m=-\infty}^{\infty}(|m\rangle\langle g|+|g\rangle\langle m|) \nonumber \\
&&
=
\Delta_{g}|g\rangle\langle g|
+
\sum_{m=-\infty}^{\infty}(m-n)\Delta|m\rangle\langle m|
+
W\sum_{m=-\infty}^{\infty}(|m\rangle\langle g|+|g\rangle\langle m|) \nonumber \\
&&\quad
+
n\Delta\hat{\mbox{\boldmath $I$}}.
\label{transformed-Hamiltonian-1}
\end{eqnarray}
In the above derivation,
we use the fact that the identity operator is given by
\begin{equation}
\hat{\mbox{\boldmath $I$}}
=
|g\rangle\langle g|
+
\sum_{m=-\infty}^{\infty}
|m\rangle\langle m|.
\end{equation}

We pay attention to the fact that
a weight $m\Delta$ of a summation for an operator $|m\rangle\langle m|$ is replaced with a new weight $(m-n)\Delta$
in the second term of Equation~(\ref{transformed-Hamiltonian-1}).
This suggests the following.
We describe new expansion coefficients of $|g\rangle$ and $\{|m\rangle\}$
as $b'(t)$ and $\{c_{m}'(t)\}$ respectively after
the transformation $\Delta_{g}\to\Delta_{g}+n\Delta$.
Then, the expansion coefficients change as
\begin{equation}
b(t)\to b'(t)=e^{-in\Delta t}b(t),
\end{equation}
\begin{equation}
c_{m}(t)
\to
c_{m}'(t)=e^{-in\Delta t}c_{m-n}(t)\quad \mbox{for $m=0, \pm 1, \pm 2, ...$.}
\end{equation}

Thus, the transformation $\Delta_{g}\to\Delta_{g}+n\Delta$
lets $|b(t)|^{2}$ and $\sum_{m=-\infty}^{\infty}|c_{m}(t)|^{2}$ be invariant.
In other words,
$|b'(t)|^{2}=|b(t)|^{2}$
and
$\sum_{m=-\infty}^{\infty}|c_{m}'(t)|^{2}=\sum_{m=-\infty}^{\infty}|c_{m}(t)|^{2}$
hold.
Hence, from Equations~(\ref{K3-definition-0}), (\ref{K3dash-definition-0}), (\ref{C21-form-1}), (\ref{C31-form-1}), and (\ref{C32-form-1}),
$K_{3}$ and $K_{3}'$ are invariant.

Second,
we prove that $K_{3}$ and $K_{3}'$ are invariant under a transformation $\Delta_{g}\to -\Delta_{g}$.
Here, we consider replacement of the Hamiltonian $\hat{H}$ with $-\hat{H}$.
This act causes transformations
$\Delta_{g}\to -\Delta_{g}$,
$\Delta\to -\Delta$,
and
$W\to -W$.
It implies the transformation of time reversal.
That is to say,
the time evolution operator is transformed as
$\exp(-i\hat{H}t)\to\exp(i\hat{H}t)$.
To compute $C_{21}$, $C_{31}$, and $C_{32}$,
we need $|b(1,\{0\};\tau)|^{2}$ and $|b(1,\{0\};2\tau)|^{2}$.
If we put an initial state with $b(0)=1$ and $c_{n}(0)=0$ for $n=0, \pm 1, \pm 2, ...$,
$b(1,\{0\};t)$ obtained by the Hamiltonian $\hat{H}$
and
$b'(1,\{0\};t)$ obtained by the Hamiltonian $-\hat{H}$
are connected as
\begin{equation}
b'(1,\{0\};t)
=
b(1,\{0\};t)^{*}.
\end{equation}
This is because the time evolution operator of $b'(1,\{0\};t)$
is the Hermitian conjugate of the time evolution operator of $b(1,\{0\};t)$.
Thus, the following equation holds:
\begin{equation}
|b'(1,\{0\};t)|^{2}
=
|b(1,\{0\};t)|^{2},
\end{equation}

To compute $C_{32}$, we need $|b(0,\{\tilde{c}_{m}\};\tau)|^{2}$.
The definition of $\tilde{c}_{n}$ is given by Equation~(\ref{definition-c-tilde-n}).
The replacement of the Hamiltonian $\hat{H}$ with $-\hat{H}$
clearly makes a transformation $\tilde{c}_{m}\to \tilde{c}_{m}'=\tilde{c}_{m}^{*}$ happen.
Thus, in the time evolution caused by the Hamiltonian $-\hat{H}$,
we obtain
\begin{equation}
b'(0,\{\tilde{c}_{m}'\};\tau)
=
b(0,\{\tilde{c}_{m}\};\tau)^{*}.
\end{equation}
Thus, the following equation holds:
\begin{equation}
|b'(0,\{\tilde{c}_{m}'\};\tau)|^{2}
=
|b(0,\{\tilde{c}_{m}\};\tau)|^{2}.
\end{equation}
Hence, for the transformations
$\Delta_{g}\to -\Delta_{g}$,
$\Delta\to -\Delta$, and
$W\to -W$,
$K_{3}$ and $K_{3}'$ are invariant.

Next,
we consider a transformation $W\to -W$
in the Hamiltonian given by Equation~(\ref{Bixon-Jortner-Hamiltonian-0}).
We concentrate on the following two cases:
\begin{description}
\item[Case 1.] We put $b(0)=1$ and $c_{n}(0)=0$ for $n=0, \pm 1, \pm 2, ...$.
\item[Case 2.] We put $b(0)=0$ and let $c_{n}(0)$ be equal to an arbitrary complex value for $n=0, \pm 1, \pm 2, ...$.
\end{description}

In Case 1, because of Equations~(\ref{explicit-form-b-1}) and (\ref{explicit-form-c-0}),
we obtain
\begin{equation}
b(t)\to b'(t)=b(t),
\end{equation}
\begin{equation}
c_{n}(t)\to c_{n}'(t)=-c_{n}(t) \quad \mbox{for $n=0, \pm 1, \pm 2, ...$}.
\end{equation}
By contrast, in Case 2,
we obtain
\begin{equation}
b(t)\to b'(t)=-b(t),
\end{equation}
\begin{equation}
c_{n}(t)\to c_{n}'(t)=c_{n}(t) \quad \mbox{for $n=0, \pm 1, \pm 2, ...$}.
\end{equation}
Thus,
for the transformation $W\to -W$,
$|b(t)|^{2}$ and $|c_{n}(t)|^{2}$ for $n=0, \pm 1, \pm 2, ...$
are invariant in both Cases 1 and 2.
In other words,
$|b'(t)|^{2}=|b(t)|^{2}$
and
$|c_{n}'(t)|^{2}=|c_{n}(t)|^{2}$ for $n=0, \pm 1, \pm 2, ...$ hold
under the transformation $W\to -W$ in both Cases 1 and 2.

From Equations~(\ref{C21-form-1}) and (\ref{C31-form-1}),
in order to compute $C_{21}$ and $C_{31}$,
we let an initial state be given by $b(0)=1$ and $c_{n}(0)=0$ for $n=0, \pm 1, \pm 2, ...$.
This corresponds with Case 1.
From Equation~(\ref{C32-form-1}),
in order to compute $C_{32}$,
we consider two initial states.
The first one is given by $b(0)=1$ and $c_{n}(0)=0$ for $n=0, \pm 1, \pm 2, ...$.
The second one is putting $b(0)=0$ and letting $c_{n}(0)$ be equal to $\tilde{c}_{n}$
for $n=0, \pm 1, \pm 2, ...$.
These initial states correspond with Cases 1 and 2, respectively.
Thus, $K_{3}$ and $K_{3}'$ are invariant
under the transformation $W\to -W$.

Putting the two transformations $\hat{H}\to -\hat{H}$ and $W\to -W$ together,
we obtain a new couple of transformations $\Delta_{g}\to -\Delta_{g}$ and $\Delta\to -\Delta$.
However,
the transformation $\Delta\to -\Delta$ causes only replacement of the expansion coefficient $c_{n}(t)$ with $c_{-n}(t)$ for the ket vector $|n\rangle$
and does not affect computations of $K_{3}$ and $K_{3}'$.
Thus, under the transformation $\Delta_{g}\to -\Delta_{g}$,
$K_{3}$ and $K_{3}'$ are invariant.

From the above considerations,
we conclude that $K_{3}$ and $K_{3}'$ are invariant under the transformations
$\Delta_{g}\to\pm\Delta_{g}+n\Delta$ for $n=0, \pm 1, \pm 2, ...$
with constant values $\Delta$ and $W$.
Similarly,
we can show that $C_{\mbox{\scriptsize rel.ent.}}$
defined in Equation~(\ref{C_rel_ent_formula-1}) is invariant under these transformations.

\begin{figure}
\begin{center}
\includegraphics[scale=1.0]{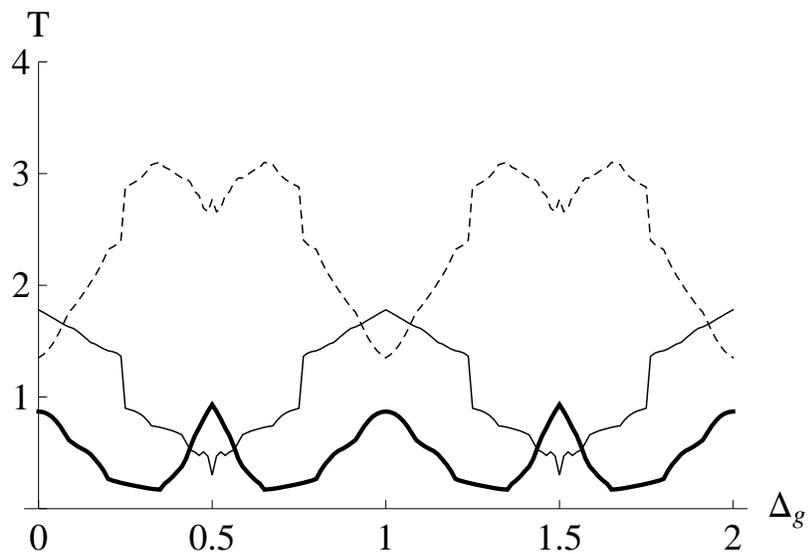}
\end{center}
\caption{Graphs of a sum of intervals of time during which $K_{3}>1$ holds,
a sum of intervals of time during which $K_{3}'>1$ holds,
and a sum of intervals of time during which $K_{3}\leq 1$ and $K_{3}'\leq 1$ hold for $0\leq T\leq 4$.
A variable of the horizontal axis is $\Delta_{g}$.
We plot the graphs in the range of $0\leq \Delta_{g}\leq 2.0$ with putting $\Delta=1.0$ and $W=0.4$.
A thick solid curve and a thin solid curve represent the cases with $K_{3}>1$ and $K_{3}'>1$, respectively.
A thin dashed curve represent the case with $K_{3}\leq 1$ and $K_{3}'\leq 1$.
Because the thin dashed curve is always above the horizontal line $T=0$,
we can consider $K_{3}$ and $K_{3}'$ not to be complementary.}
\label{Figure06}
\end{figure}

Figure~\ref{Figure06} shows graphs of a sum of intervals of time during which $K_{3}>1$ holds,
a sum of intervals of time during which $K_{3}'>1$ holds,
and a sum of intervals of time during which $K_{3}\leq 1$ and $K_{3}'\leq 1$ hold for $0\leq T\leq 4$.
A variable of the horizontal axis is $\Delta_{g}$.
Looking at Figure~\ref{Figure06},
we can confirm that the time evolution of $K_{3}$ and $K_{3}'$ is invariant under
the transformations $\Delta_{g}\to\pm\Delta_{g}+n\Delta$ for $n=0, \pm 1, \pm 2, ...$.

Here, we pay attention to the fact
that both $K_{3}>1$ and $K_{3}'>1$ never hold at the same time.
The reason why is as follows.
If we assume $K_{3}>1$ and $K_{3}'>1$ simultaneously,
we obtain
\begin{eqnarray}
C_{21}+C_{32}-C_{31}
&>&
1, \nonumber \\
-C_{21}-C_{32}-C_{31}
&>&
1.
\end{eqnarray}
Then, we have
$-C_{31}>1$.
However, because of the definition of $C_{31}$,
$-1\leq C_{31}\leq 1$ has to be satisfied by any measurement.
Thus,
both $K_{3}>1$ and $K_{3}'>1$ never hold at the same time.

\begin{figure}
\begin{center}
\includegraphics[scale=1.0]{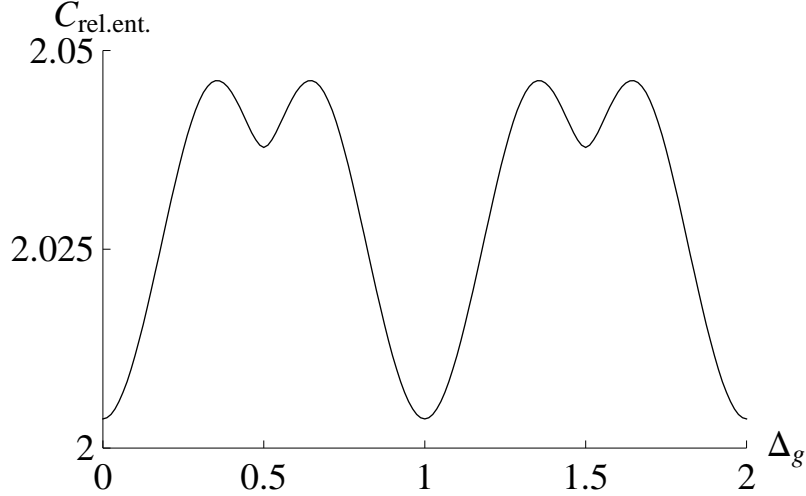}
\end{center}
\caption{A graph of an average of $C_{\mbox{\scriptsize rel.ent.}}$ for $0\leq T\leq 4$.
A variable of the horizontal axis is $\Delta_{g}$.
We plot the graph in the range of $0\leq\Delta_{g}\leq 2.0$
with putting $\Delta=1.0$ and $W=0.4$.
Although the average of $C_{\mbox{\scriptsize rel.ent.}}$ is not sensitive to the variation of $\Delta_{g}$,
it shows the invariance under the transformation of $\Delta_{g}$.}
\label{Figure07}
\end{figure}

Figure~\ref{Figure07} shows a graph of an average of $C_{\mbox{\scriptsize rel.ent.}}$ for $0\leq T\leq 4$.
A variable of the horizontal axis is $\Delta_{g}$.
Looking at Figure~\ref{Figure07},
we can confirm that the time evolution of $C_{\mbox{\scriptsize rel.ent.}}$ is invariant
under the transformations $\Delta_{g}\to\pm\Delta_{g}+n\Delta$ for $n=0, \pm 1, \pm 2, ...$.

To calculate $K_{3}$, $K_{3}'$, and $C_{\mbox{\scriptsize rel.ent.}}$ numerically,
we use the Fortran compiler with the quadruple precision for real values.
In actual calculations,
instead of infinite dimensional space,
we assume that the dimension of the Hilbert space is finite and its orthogonal basis is given by
$|g\rangle$ and $\{|n\rangle:n=0, \pm 1, \pm 2, ..., \pm 1200\}$.

\section{\label{section-discussion}Discussion}
So far, only the special solution has been known for the Bixon-Jortner model.
In this paper, we analytically derive the general solution of the model.
This new rigorous solution lets us be able to compute the Leggett-Garg inequalities and the relative entropy of coherence exactly.
This fact is one of novel points in this paper.

Turning our eyes to Figures~\ref{Figure03}, \ref{Figure04}, and \ref{Figure06},
we become aware that there are temporal intervals during which $K_{3}\leq 1$ and $K_{3}'\leq 1$ hold.
This indicates that $K_{3}$ and $K_{3}'$ are not complementary.
By contrast,
Friedenberger and Lutz chose an observable and let
$K_{3}$ and $K_{3}'$ be complementary for an isolated two-level system in Reference~\cite{Friedenberger2017}.
Moreover, they showed that $K_{3}$ and $K_{3}'$ are not complementary in a damped two-level system.
In the present paper, we consider a pure state that develops according to the Hamiltonian given by Equation~(\ref{Bixon-Jortner-Hamiltonian-0}).
Although the system does not suffer from decoherence,
$K_{3}$ and $K_{3}'$ are not complementary.
We cannot find an proper observable which lets $K_{3}$ and $K_{3}'$ be complementary.

When we obtain $K_{3}\leq 1$ and $K_{3}'\leq 1$ during a temporal interval,
we consider that the system may lose quantumness.
Thus, we can expect the relative entropy of coherence to be suppressed in this interval.
However, we cannot observe such a phenomenon.

Although the Hamiltonian of the Bixon-Jortner model is simple,
its analytical solution is very complicated.
One of features the exact solution has is that
we can find kicks in the time evolution.
The authors of current paper think that we can extract more interesting characteristics from the Bixon-Jortner model in the near future.

\section*{Acknowledgement}
The authors thank Naomichi Hatano for careful and critical reading of the manuscript of the current paper.


\begin{thebibliography}{99}
%
\bibitem{Bixon1968}
M.~Bixon and J.~Jortner,
`Intramolecular radiationless transitions',
J. Chem. Phys. {\bf 48}(2), 715--726 (1968).
doi:10.1063/1.1668703
%
\bibitem{Englman1970}
R.~Englman and J.~Jortner,
`The energy gap law for radiationless transitions in large molecules',
Mol. Phys. {\bf 18}(2), 145--164 (1970).
doi:10.1080/00268977000100171
%
\bibitem{Jortner1976}
J.~Jortner,
`Temperature dependent activation energy for electron transfer between biological molecules',
J. Chem. Phys. {\bf 64}(12), 4860--4867 (1976).
doi:10.1063/1.432142
%
\bibitem{Stey1972}
G.C.~Stey and R.W.~Gibberd,
`Decay of quantum states in some exactly soluble models',
Physica {\bf 60}(1), 1--26 (1972).
doi:10.1016/0031-8914(72)90218-2
%
\bibitem{Lefebvre1974}
R.~Lefebvre and J.~Savolainen,
`Memory functions and recurrences in intramolecular processes',
J. Chem. Phys. {\bf 60}(6), 2509--2515 (1974).
doi:10.1063/1.1681390
%
\bibitem{Yeh1982}
J.J.~Yeh, C.M.~Bowden, and J.H.~Eberly,
`Interrupted coarse-grained theory of unimolecular relaxation and stimulated recurrences
in photoexcitation of a quasicontinuum',
J. Chem. Phys. {\bf 76}(12), 5936--5946 (1982).
doi:10.1063/1.442948
%
\bibitem{Milonni1983}
P.W.~Milonni, J.R.~Ackerhalt, H.W.~Galbraith, and M.-L.~Shih,
`Exponential decay, recurrences, and quantum-mechanical spreading in a quasicontinuum model',
Phys. Rev. A {\bf 28}(1), 32--39 (1983).
doi:10.1103/PhysRevA.28.32
%
\bibitem{Skinner1981}
J.L.~Skinner, H.C.~Andersen, and M.D.~Fayer,
`Correlation-function analysis of coherent optical transients and fluorescence from a quasi-two-level system',
Phys. Rev. A {\bf 24}(4), 1994--2008 (1981).
doi:10.1103/PhysRevA.24.1994
%
\bibitem{Bar-Ad1997}
S.~Bar-Ad, P.~Kner, M.V.~Marquezini, S.~Mukamel, and D.S.~Chemla,
`Quantum confined Fano interference',
Phys. Rev. Lett. {\bf 78}(7), 1363--1366 (1997).
\\
doi:10.1103/PhysRevLett.78.1363
%
\bibitem{Santra2017}
S.~Santra, B.~Cruikshank, R.~Balu, and K.~Jacobs,
`Fermi's golden rule, the origin and breakdown of Markovian master equations, and the relationship between oscillator baths and the random matrix model',
J. Phys. A: Math. Theor. {\bf 50}(41), 415302 (2017).
doi:10.1088/1751-8121/aa8777
%
\bibitem{Radmore1987}
P.M.~Radmore, S.~Tarzi, and P.L.~Knight,
`Rates and recurrences in quasicontinuum photoexcitation',
J. Mod. Opt. {\bf 34}(5), 587--606 (1987).
\\
doi:10.1080/09500348714550621
%
\bibitem{Tarzi1988}
S.~Tarzi and P.M.~Radmore,
`Dressed states and spectra in quasicontinuum excitation',
Phys. Rev. A {\bf 37}(12), 4734--4740 (1988).
doi:10.1103/PhysRevA.37.4734
%
\bibitem{Tarzi1989}
S.~Tarzi, P.M.~Radmore, and S.M.~Barnett,
`Sparse and trapping dressed states in a quasicontinuum system',
J. Phys. B: At. Mol. Opt. Phys. {\bf 22}(18), 2935--2940 (1989).
doi:10.1088/0953-4075/22/18/015
%
\bibitem{Radmore1995}
P.M.~Radmore,
`Photoexcitation to a periodic continuum',
J. Mod. Opt. {\bf 42}(3), 579--584 (1995).
doi:10.1080/09500349514550541
%
\bibitem{Barnett1997}
S.M.~Barnett
and
P.M.~Radmore,
{\it Methods in Theoretical Quantum Optics}
(Oxford University Press, Oxford, 1997).
%
\bibitem{Leggett1985}
A.J.~Leggett and A.~Garg,
`Quantum mechanics versus macroscopic realism: Is the flux there when nobody looks?',
Phys. Rev. Lett. {\bf 54}(9), 857--860 (1985).
doi:10.1103/PhysRevLett.54.857
%
\bibitem{Emary2014}
C.~Emary, N.~Lambert, and F.~Nori,
`Leggett-Garg inequalities',
Rep. Prog. Phys. {\bf 77}(1), 016001 (2014).
doi:10.1088/0034-4885/77/1/016001
%
\bibitem{Palacios-Laloy2010}
A.~Palacios-Laloy, F.~Mallet, F.~Nguyen, P.~Bertet, D.~Vion, D.~Esteve, and A.N.~Korotkov,
`Experimental violation of a Bell's inequality in time with weak measurement',
Nat. Phys. {\bf 6}(6), 442--447 (2010).
doi:10.1038/nphys1641
%
\bibitem{Goggin2011}
M.E.~Goggin, M.P.~Almeida, M.~Barbieri, B.P.~Lanyon, J.L.~O'Brien, A.G.~White, and G.J.~Pryde,
`Violation of the Leggett-Garg inequality with weak measurements of photons',
Proc. Natl. Acad. Sci. U.S.A. {\bf 108}(4), 1256-1261 (2011).
\\
doi:10.1073/pnas.1005774108
%
\bibitem{Knee2012}
G.C.~Knee, S.~Simmons, E.M.~Gauger, J.J.L.~Morton, H.~Riemann, N.V.~Abrosimov, P.~Becker, H.-J.~Pohl, K.M.~Itoh, M.L.W.~Thewalt,
G.A.D.~Briggs, and S.C.~Benjamin,
`Violation of a Leggett-Garg inequality with ideal non-invasive measurements',
Nat. Commun. {\bf 3}, 606 (2012).
doi:10.1038/ncomms1614
%
\bibitem{Baumgratz2014}
T.~Baumgratz, M.~Cramer, and M.B.~Plenio,
`Quantifying coherence',
Phys. Rev. Lett. {\bf 113}(14), 140401 (2014).
doi:10.1103/PhysRevLett.113.140401
%
\bibitem{Zhang2016}
Y.-R.~Zhang, L.-H.~Shao, Y.~Li, and H.~Fan,
`Quantifying coherence in infinite-dimensional systems',
Phys. Rev. A {\bf 93}(1), 012334 (2016).
\\
doi:10.1103/PhysRevA.93.012334
%
\bibitem{Friedenberger2017}
A.~Friedenberger and E.~Lutz,
`Assessing the quantumness of a damped two-level system',
Phys. Rev. A {\bf 95}(2) 022101 (2017).
doi:10.1103/PhysRevA.95.022101
%
\end{thebibliography}
\end{document}